\documentclass[12pt]{aastex}

\newcommand{\EQ}{\begin{equation}}
\newcommand{\EN}{\end{equation}}
\newcommand{\EQA}{\begin{eqnarray}}
\newcommand{\ENA}{\end{eqnarray}}
\usepackage{amsmath}
\newcommand{\A}{\mathcal{A}}

\newcommand{\B}{\mathcal{B}}
\newcommand{\R}{\mathcal{R}}

\newcommand{\uu}{{\bf u}}
\newcommand{\bb}{{\bf b}}
\newcommand{\BB}{{\bf B}}

\newcommand{\UU}{{\bf U}}

\begin{document}

\title{Dynamo efficiency with shear in helical turbulence} 

\author{Nicolas Leprovost \and Eun-jin Kim}

\affil{Department of Applied Mathematics, University of Sheffield, Sheffield S3 7RH, UK}

\begin{abstract}
To elucidate the influence of shear flow on the generation of magnetic field (the dynamo effect), we study the kinematic limit where the magnetic field does not backreact on the velocity field. By non-perturbatively incorporating the effect of shear in a helically forced turbulence, we show that turbulence intensity and turbulent transport coefficients (turbulent viscosity, $\alpha$ and $\beta$ effect) are enhanced by a weak shear while strongly suppressed for strong shear. In particular, $\beta$ is shown to be much more strongly suppressed than $\alpha$ effect. We discuss its important implications for dynamo efficiency, i.e. on the scaling of the dynamo number with differential rotation.
\end{abstract}


It is now widely accepted that astrophysical and geophysical magnetic fields are not the remains of a fossil field created during the formation of planets or stars (as they would have decayed on a time-scale much shorter than their current life time), but are self-excited by motions of conductive fluid (for instance, molten iron within the outer liquid core for the Earth and conducting plasma for the Sun). The evolution of a magnetic field ${\bf B}$ in a conducting fluid ${\bf V}$ is governed by the induction equation:
\EQ
\partial_t {\bf B} + {\bf V} \cdot {\bf \nabla} \BB =  \BB \cdot {\bf \nabla} \, {\bf V} + \eta \nabla^2 {\bf B} \quad  \text{and} \quad {\bf \nabla \cdot \BB} = 0 \; , \label{Induction}
\EN
where $\eta$ is the ohmic diffusivity. The first term on the right hand side (RHS) of Eq. (\ref{Induction}) is the  stretching of magnetic field lines by gradients of the velocity field.

While laminar flows that can generate magnetic fields (dynamo) have been known for a long time, the effect of turbulence on the generation of large-scale coherent magnetic field remains controversial. A main problem is that turbulence tends to create magnetic field at small scales (i.e. scale comparable to the original velocity field) while observations of astrophysical magnetic fields (for instance galaxies) reveal coherent magnetic field on a scale much larger than the fluctuating velocity field. Theories, such as mean-field dynamo \citep{Moffatt78,Krause80}, have investigated the necessary ingredients for large-scale field generation. In the framework of mean-field dynamo, the magnetic and velocity fields can be decomposed into mean and fluctuating parts: ${\bf B} = \langle {\bf B} \rangle + {\bf b}$  and ${\bf V} = \langle {\bf V} \rangle + {\bf v}$, where the $\langle \bullet \rangle$ stands for an average on the realization of the small-scale fields. Substitution of this decomposition into Eq. (\ref{Induction}) and averaging yield the following equation for the mean magnetic field:
\EQ
\label{Induction2}
\partial_t \langle {\bf B} \rangle + \langle {\bf V} \rangle \cdot {\bf \nabla} \langle \BB \rangle =  \langle \BB \rangle \cdot {\bf \nabla} \, \langle {\bf V} \rangle + \eta \nabla^2 \langle {\bf B} \rangle + {\bf \nabla} \times {\bf \mathcal{E}} \; .
\EN
The first term on the RHS of Eq. (\ref{Induction}) represents the stretching of magnetic field lines by gradient of the mean flow (${\bf \nabla} \, \langle {\bf V}\rangle$) and is called the $\Omega$ effect. It is an efficient mechanism to create toroidal field from a poloidal field in a system with differential rotation \citep{Moffatt78}. The term ${\bf \mathcal{E}} = \langle {\bf v} \times {\bf b} \rangle$ is the electromotive force, which is often taken to be linear in the mean magnetic field ($\langle {\bf B} \rangle$). In the case of an isotropic turbulence, this can be simplified as:
\EQ
\label{Electromotive}
\mathcal{E}_i = \alpha \langle B_i \rangle - \beta  ({\bf \nabla \times \langle B \rangle})_i + \dots  \; .
\EN
The structure of the electromotive force permits the possibility of other mechanisms for the amplification of the large-scale magnetic field besides the $\Omega$ effect. The one that has been discussed most is the $\alpha$ effect, the first term on the RHS of Eq. (\ref{Electromotive}). This has been shown to generate magnetic field at large scale for a helical turbulence. Thus, it is a perfect candidate to explain magnetic fields in systems influenced by Coriolis force (which produces a net helicity) such as in stellar convection zones. This type of dynamo is thus classified as $\alpha\Omega$ if the $\Omega$ effect (measured by the strength of the shear $\Omega$ in our notations) is stronger than the $\alpha$ effect, or $\alpha^2$ type if the $\alpha$ effect dominates over the $\Omega$ effect. The second term in the RHS is Eq. (\ref{Electromotive}) is the turbulent diffusivity which adds up to the molecular diffusivity $\eta$. Consequently, if $\beta$ is positive, it inhibits the growth of magnetic field.

Recently, numerical simulations have shown dynamo action at large scale in non-helical turbulence in the presence of shear \citep{Yousef08}. This is an interesting result as the $\alpha$ effect is often thought to vanish in a turbulence without helicity. Various mechanisms have been invoked to explain this large-scale dynamo: stochastic $\alpha$ effect \citep{Proctor07}, shear amplification of small-scale dynamo \citep{Blackman98}, magnetic effect driven by current helicity flux \citep{Vishniac01} or negative diffusivity \citep{Urpin02}. Another possibility is the shear current effect \citep{Rogachevskii03} which appears in a turbulent flow with a mean shear flow. In that case, the expression of the $\beta$ coefficient can be rewritten $\beta_{ijk} = - \beta^T \epsilon_{ijk} + F_{ijk}({\bf \nabla} \UU_0)$ where $\beta^T$ is the turbulent magnetic diffusion while the second term proportional to shear ${\bf \nabla} \UU_0$ acts as a source of magnetic field \citep{Rogachevskii03}. It is thus of prime importance to investigate how the electromotive force (and consequently the $\alpha$ and $\beta$ coefficients) depends on a large-scale shear flow \citep{Rogachevskii03,Rogachevskii04,Radler06,Brandenburg08}. In all these previous studies, strong shear is conductive to dynamo as it creates magnetic energy via the $\Omega$ effect, acts as a source of magnetic field (e.g. via the shear-current effect), causes instability \citep{Tobias04}, etc. 

\begin{figure}
\begin{center}
\includegraphics[scale=0.5,clip]{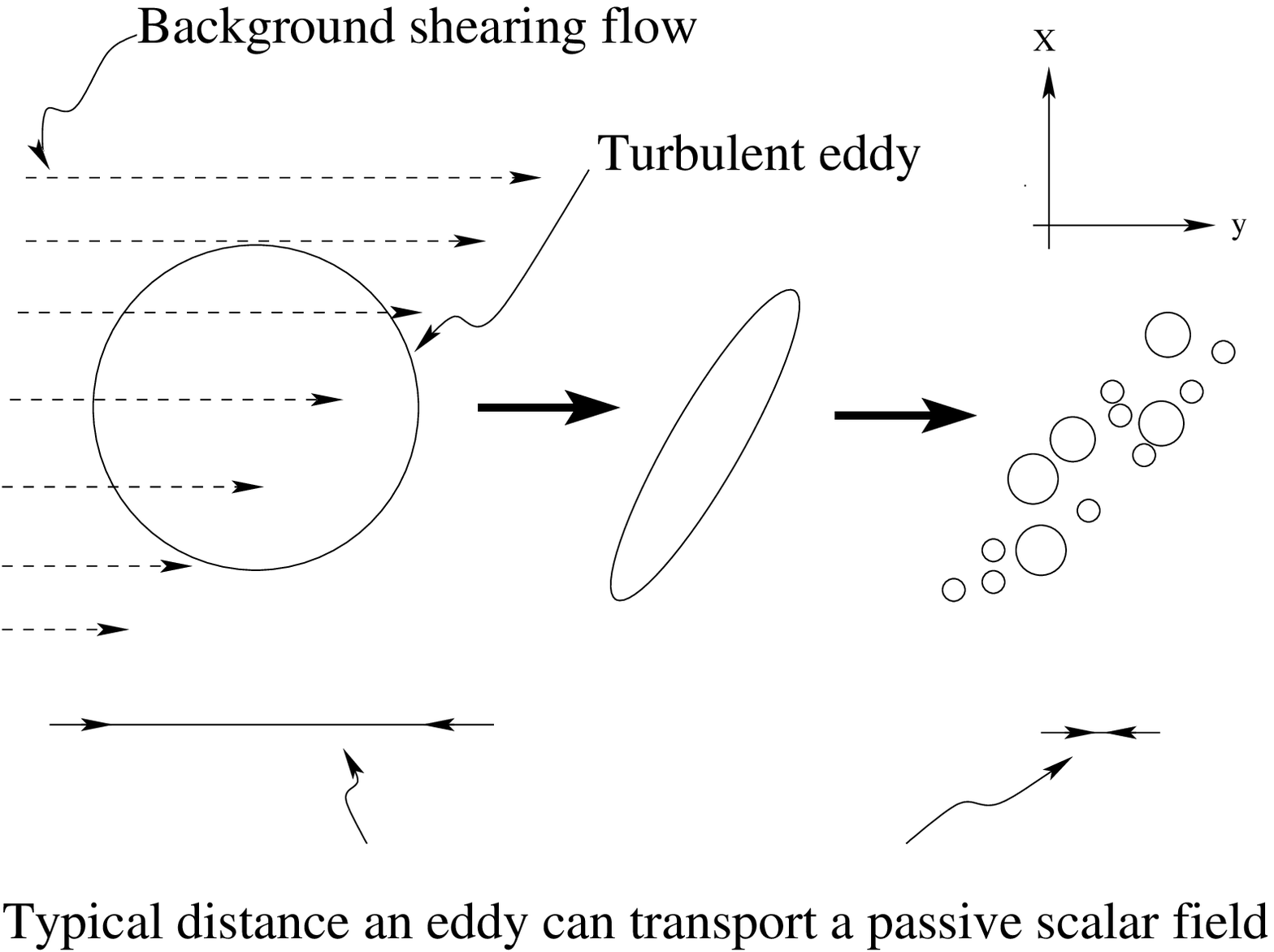}
\caption{\label{ShearEff} Sketch of the effect of shear on a turbulent eddy.}
\end{center}
\end{figure}

One interesting problem, which has not been investigated by most previous authors, is the effect of a stable shear flow on turbulent transport through the modification of the properties of turbulence alone, without direct influence on $\langle {\bf B} \rangle$ (i.e. no $\Omega$-effect, shear-current effect). A strong shear flow, without altering $\langle {\bf B} \rangle$ directly, can reduce turbulent transport as turbulence becomes weak by shear stabilization \citep{Burrell97}. This is basically because shear advects turbulent eddies differentially, elongating and distorting their shapes, thereby rapidly generating small scales which are ultimately disrupted by molecular dissipation on small scales (see Fig. \ref{ShearEff}). As a result, turbulence level as well as turbulent transport of various quantities can be significantly reduced compared to the case without shear \citep{Kim05,Kim06,2Shears}. In particular, in the case when a stable shear flow is parallel to the magnetic field, a dramatic quenching of turbulent magnetic diffusion ($\beta$-effect) was clearly shown in a recent numerical simulation of 2D MHD turbulence \citep{Newton08}. In 3D MHD turbulence, by considering a stable shear flow parallel to a uniform large-scale magnetic field, \citet{Quenching} theoretically predicted that the $\alpha$ effect is quenched by shear as well as magnetic field. In particular, in the kinematic case (for weak magnetic field), the $\alpha$ effect was shown to be reduced as flow shear $\A$ increases with the scaling $\A^{-5/3}$. However, to understand fully the effect of shear on the dynamo process, it remains to compute its effect on the turbulence diffusion of magnetic field, i.e. the $\beta$ effect, by considering a non-uniform magnetic field. This is what we do in the remainder of this letter.

In the kinematic limit, the backreaction of the magnetic field on the velocity is neglected. From the physical point of view, this amounts to considering a very weak magnetic field and ignoring the Lorentz Force on the fluid which is quadratic in the magnetic field.  For an incompressible conducting fluid, the resulting equations of motion are:
\EQA
\label{eqMHD}
\partial_t {\bf V} + {\bf V} \cdot {\bf \nabla} {\bf V} &=& - {\bf \nabla} p + \nu \Delta {\bf V} + {\bf f} \; , \\ \nonumber
\partial_t \BB + {\bf V} \cdot {\bf \nabla} \BB &=&  \BB \cdot {\bf \nabla} {\bf V} + \eta \Delta \BB  \; , \\ \nonumber
{\bf \nabla} \cdot {\bf V} &=& {\bf \nabla} \cdot \BB = 0 \; .
\ENA
Here $B$ is the Alfv\'en speed, $p$ is the total (hydrodynamical + magnetic) pressure and ${\bf f}$ is a small-scale forcing. To study the effect of shear flows and magnetic fields on small-scale turbulence, we prescribe a large scale flow of the form $\UU_0 = - x \A {\bf e_y}$ and a sheared large-scale magnetic field ${\bf B}_0 = (B_0 - \B x) {\bf e_y}$. $\BB_0$ has been chosen parallel to $\UU_0$ so that there is no direct interaction between the two fields, e.g. excluding the $\Omega$-effect in our study (in contrast with the case considered by \citep{Yousef08,Schekochihin08}, etc.). To solve the equations for the fluctuating velocity field, $\uu = {\bf V} - \UU_0$, and magnetic field, $\bb = \BB - \BB_0$, we use the quasi-linear approximation assuming that the interaction between fluctuating fields is negligible compared to the interaction between large and small-scale fields. The equations for the fluctuating fields can then be written as:
\EQA
\label{eqRDT}
\partial_t \uu + \uu \cdot {\bf \nabla} \UU_0  &=& - {\bf \nabla} p  + \nu \Delta \uu + {\bf f} \; , \\ \nonumber
\partial_t \bb + \uu \cdot {\bf \nabla} \BB_0 + \UU_0 \cdot {\bf \nabla} \bb &=&  \bb \cdot {\bf \nabla} \UU_0 + \BB_0 \cdot {\bf \nabla} \uu + \eta \Delta \bb  \; , \\ \nonumber
{\bf \nabla} \cdot \uu &=& {\bf \nabla} \cdot \bb = 0 \; .
\ENA
In the sequel, we shall assume an unit magnetic Prandtl number ($\nu = \eta$) and introduce a time dependent Fourier transform \citep{Kim05}:
\EQ
\nonumber
Y(\vec{x},t) = \frac{1}{(2\pi)^3} \int d^3 \vec{k} e^{i\bigl[{ k_x(t)} x + k_y y + k_z z\bigr]}
\tilde{Y}(\vec{k},t) \; .
\EN
Transforming the time variable from $t$ to $\tau = k_x(t) / k_y = k_x(t_0) / k_y + \A (t-t_0)$, Eq. (\ref{eqRDT}) can be written:
\EQA
\label{eqRDT2}
\partial_\tau \tilde{u}_i - \tilde{u}_x \, \delta_{i2} &=& - i k_y \theta_i \tilde{p} - \xi (g^2+\tau^2) \tilde{u}_i + \tilde{f}_i \; , \\ \nonumber
\partial_\tau \tilde{b}_i - \R \tilde{u}_x \delta_{i2} &=& - \tilde{b}_x \delta_{i2} + \R \partial_\tau \tilde{u}_i + i \gamma \tilde{u}_i - \xi (g^2+\tau^2) \tilde{b}_i  \; , \\ \nonumber
\tau \tilde{u}_x + \tilde{u}_y + \beta \tilde{u}_z &=& \tau \tilde{b}_x + \tilde{b}_y + \beta \tilde{b}_z = 0 \; .
\ENA
Here, $\R = \B / \A$ and $\gamma = B_0 k_y /\A$ are the ratio of the magnetic shear and constant magnetic field to the velocity shear, respectively; $\beta = k_z / k_y$ and $g^2=1+\beta^2$; $\xi=\nu k_y^2 / \A$ and $\theta_i = (\tau,1,\beta)$. Note that since the first equation of (\ref{eqRDT2}) does not involve the magnetic field, the solution to $v_i$ is the same as in the hydrodynamical case \citep{Kim05}. Using the velocity from \citet{Kim05}, the magnetic fluctuations can be obtained from the second equation of (\ref{eqRDT2})  as:
\EQA
\label{Magnetic}
\tilde{b}_x &=& \int_{\tau_0}^{\tau} dt \frac{f_x(t)(g^2+t^2)}{\A} e^{G(t,\tau)}  \left[\frac{\R}{g^2+\tau^2} + i \gamma  \{ T(\tau) - T(t) \} - \R \xi (\tau-t) \right] \; , \\ \nonumber
\tilde{b}_z &=& \int_{\tau_0}^{\tau} dt \frac{f_z(t)}{\A}  e^{G(t,\tau)}  \left[\R (1-\xi \{Q(\tau)-Q(t)\}) + i \gamma (\tau - t) \right] \\ \nonumber
&& - \beta \int_{\tau_0}^{\tau} dt  \frac{f_x(t)(g^2+t^2)}{\A } e^{G(t,\tau)}
\left[\R \{I(t,\tau)- \xi J_2\} + i \gamma J_1  \right]  \; .
\ENA
Here,
\EQA
\label{Ittau}
G(t,\tau) &=& -\xi \{Q(\tau)-Q(t)\} \; , \qquad Q(x)=g^2 x + x^3/3 \; , \\ \nonumber
I(t,\tau) &=& \frac{1}{2g^2} \left[\frac{\tau}{g^2+\tau^2} - \frac{t}{g^2+t^2} + T(\tau) - T(t) \right] \; , \qquad  \\ \nonumber
J_1 &=& \int_t^{\tau} I(t,x) \, dx \; ,  \; \text{and} \; J_2 = \int_t^{\tau} I(t,x) (g^2+x^2) \,  dx \; ,
\ENA
where $T(x) = \arctan(x/g) / g$. $\tilde{b}_y$ can be obtained using incompressibility: $\tilde{b}_y  = - \tau \tilde{b}_x - \beta \tilde{b}_z$.

Our main interest is in the total stress and the electromotive force, which determine the growth/decay of the large-scale velocity field and the large-scale magnetic field, respectively. First, the stress is $\mathcal{S} = \langle u_x u_y \rangle - \langle b_x b_y \rangle$. This total stress consisting of Reynolds stress $\langle u_x u_y \rangle$ and Maxwell stress $\langle b_x b_y \rangle$ gives a turbulent viscosity $\nu^T$ in Navier-Stokes equation for large-scale flows, which enhances the molecular viscosity to $\nu + \nu^T$. For the assumed shear flow $U_0 = - \A x$, the turbulent viscosity is given by $\mathcal{S} = \nu^T \A$. Second, for the magnetic field considered here, the electromotive force reduces to:
\EQA
\mathcal{E}_x &=& \langle u_y b_z - u_z b_y \rangle = \alpha_{xy} B_0  \; , \\ \nonumber
\mathcal{E}_y &=& \langle u_z b_x - u_x b_z \rangle = \alpha_{yy} B_0  \; , \\ \nonumber
\mathcal{E}_z &=& \langle u_x b_y - u_y b_x \rangle = - \beta \B \; .
\ENA
Note here that only three coefficients $\alpha_{yy}$, $\alpha_{zy}$ and $\beta$ are non-vanishing in our configuration. In particular, phenomena such  as the ${\bf \Omega \times J}$ \citep{Radler06} and shear current effects \citep{Rogachevskii03}, which have been advocated to generate magnetic field for non-helical turbulence subject to rotation and shear as noted previously, are absent here \citep[in contrast with the simulations of][]{Yousef08}. Note that a shear-current effect could be studied by using a similar analysis but assuming the large-scale magnetic field to depend on $z$ rather than $x$, which will be addressed in a future contribution.

To calculate the correlation functions involved in the transport coefficients, we consider an incompressible forcing which is spatially homogeneous and temporally short correlated with the correlation time $\tau_f$. Specifically, in Fourier space, the correlation function of the forcing is taken as:
\EQ
\label{Forcing}
\langle \tilde{f}_i({\bf k_1},t_1) \tilde{f}_j({\bf k_2},t_2) \rangle = \tau_f \, (2\pi)^3 \delta({\bf k_1}+{\bf k_2}) \, \delta(t_1-t_2)  \phi_{ij}({\bf k_2}) \; ,
\EN
where the tilde denotes a Fourier-transform with respect to the spatial variable. As noted previously, the $\alpha$ effect can be linked to the helicity of the turbulent flow. Consequently, we consider a forcing with both a symmetric part (with energy spectrum $E$) and a helical part (with helicity spectrum $H$) given by:
\EQ
\phi_{lm}({\bf k}) = E(k) \left(\delta_{lm} - \frac{k_l k_m}{k^2} \right) + i \epsilon_{lmp} k_p H(k) \; . 
\EN
In the following, the turbulence intensity, turbulent viscosity and $\alpha$ effect are expressed in terms of their values in the absence of shear or magnetic field, $e_0$, $\nu_0$, $\alpha_0$ and $\beta_0$, which can be shown to be: 
\EQA
e_0 &=& \frac{\tau_f}{(2\pi)^2} \int_0^{+\infty} d k \frac{E(k)}{\nu} \; , \\ \nonumber
\nu_0 &=& \frac{\tau_f}{(2\pi)^2} \int_0^{+\infty} d k \frac{E(k)}{5 \nu^2 k^2}  \; ,  \\ \nonumber
\alpha_0 &=& - \frac{\tau_f}{(2\pi)^2} \int_0^{+\infty} d k \frac{H(k)}{6 \nu^2}  \; , \\ \nonumber
\beta_0 &=& \frac{\tau_f}{(2\pi)^2} \int_0^{+\infty} d k \frac{E(k)}{6 \nu^2 k^2}  \; .
\ENA
Using equations for velocity in \citet{Kim05} and Eq. (\ref{Magnetic}) and after a long algebra following \citet{Kim05}, we can find the turbulent intensity, stress and the electromotive force. Omitting the details, here we provide the results only for the limiting case of a  weak ($\xi = \nu k_y^2 / \A \gg 1$) and strong shear ($\xi = \nu k_y^2 / \A \ll 1$).

First, in the case where the shear is weak compared to the diffusion rate ($\xi \gg 1$), we obtain:
\EQA
\label{Coeff2}
\langle u_x^2 \rangle &\sim& \frac{2 e_0}{3} \left[1 + \frac{9 \xi_*^{-2}}{35}  \right] \; , \\ \nonumber
\langle u_z^2 \rangle &\sim& e_0 \left[1 + \frac{3 \xi_*^{-2}}{70}  \right] \; , \\ \nonumber
\langle b_x^2 \rangle &\sim& \frac{e_0}{3} \left[R^2 + \frac{\gamma^2 \xi_*^{-2}}{2} + \frac{36 \R^2 \xi_*^{-2}}{35}\right] \; , \\ \nonumber
\langle b_z^2 \rangle &\sim& \frac{e_0}{3} \left[R^2 + \frac{\gamma^2 \xi_*^{-2}}{2} + \frac{2526 \R^2 \xi_*^{-2}}{715}\right]  \; , \\ \nonumber
\nu_T &\sim& \nu_0  \left[1 + \frac{4 \xi_*^{-2}}{21}  \right]  \; , \\ \nonumber
\alpha_{xy} &\sim& \alpha_0 \frac{\xi_*^{-1}}{5}   \; , \\ \nonumber
\alpha_{yy} &\sim& \alpha_0  \left[1 + \frac{33 \xi_*^{-2}}{70}  \right]  \; , \\ \nonumber
\beta &\sim& - \beta_0 \left[1 + \frac{26 \xi_*^{-2}}{35}  \right]    \; .
\ENA 
Note that the turbulent viscosity $\nu_T$ and the $\beta$ effect are proportional only to the energy part of the forcing while the $\alpha$ effect is proportional only to the non-reflectionally symmetric part of the forcing. This is consistent with the expectation that the $\alpha$ effect is due to helical flow, which results from the helical forcing with helicity spectrum $H$. Eq. (\ref{Coeff2}) shows that (in the weak shear limit) all the turbulent coefficients increase with shear above their values without shear. The increase in $\beta$ with shear seems to be in agreement with numerical results shown in Fig. 1 of \citet{Mitra08} obtained in a slightly different configuration of ${\bf U_0}$ and ${\bf B_0}$. Eq. (\ref{Coeff2}) also shows that $\alpha_{xy} \ll \alpha_{yy}$ i.e. that the electromotive force is primarily parallel to the large-scale magnetic field (i.e. in the $y$ direction). Furthermore, without shear ($\xi^{-1}=0$), we see that $\alpha_{xy}=0$ showing that this component of the $\alpha$ effect exists only for non vanishing shear. This is due to the fact that shear induces an anisotropic turbulence \citep[see e.g.][]{RotShearAA} which in turn triggers off-diagonal components in the $\alpha$ tensor. Note that a different result was obtained by \citet{Kim01} who found in two dimensions that the turbulent diffusivity decreases with shear. This difference comes form the fact that \citet{Kim01} considered an anisotropic forcing, physically different form the isotropic forcing considered here.

In the opposite limit of strong shear ($\xi = \nu k_y^2 / \A \ll 1$), turbulence intensity and transport coefficients are obtained as follows:
\EQA
\langle u_x^2 \rangle &\sim& \xi e_0 \; , \qquad \qquad  \langle u_z^2 \rangle \sim \xi^{2/3} e_0 \; , \\ \nonumber
\langle b_x^2 \rangle &\sim& \xi^{8/3} e_0 \; , \quad \qquad \langle b_z^2 \rangle \sim \xi^{2} e_0 \; , \\ \nonumber
\nu_T &\sim& \xi^2 \nu_0 \; , \qquad  \qquad \beta \sim \xi^{7/3} \nu_0  \; , \\ \nonumber
\alpha_{xy} &\sim& \xi^{4/3} \alpha_0  \; , \quad \qquad \alpha_{yy} \sim \xi^{5/3} \alpha_0  \; .
\ENA 
These results show that in the limit of strong shear (compared to diffusion), all the turbulent quantities are reduced by shear with scalings given above. Note that the magnetic energy $\langle b^2 \rangle$ is more reduced than kinetic energy $\langle u^2 \rangle$. Furthermore, both the velocity and  magnetic field in the direction of the shear are reduced more severely than in the perpendicular direction, manifesting the anisotropic turbulence induced by shear. It is because flow shear directly influences the component parallel to itself (i.e. the $x$ component in Fig. 1) via elongation while only indirectly the other two components (i.e. the $y$ and $z$ components in Fig. 1) through enhanced dissipation. The electromotive force shows that the $x$-component of the $\alpha$ effect ($\alpha_{xy}$) is now larger than the $y$ one ($\alpha_{yy}$). This is again because, as the shear increases, the anisotropy in the flow increases enhancing the off-diagonal component $\alpha_{xy}$ strongly. Finally, the turbulent diffusivity $\beta$ is reduced as $\xi^{7/3}$ more severely than the $\alpha$ effect ($\alpha_{yy} \propto \xi^{5/3}$), which has interesting implications for the dependence of the dynamo number (characterizing the efficiency of the dynamo) with differential rotation, as discussed in the introduction.

To summarize, we found that the $\beta$ effect is reduced as $\A^{-7/3}$, with a much stronger dependence on the shear than the $\alpha$ effect ($\alpha_{yy} \propto \A^{-5/3})$. This result can have interesting implications for solar dynamo which is often envisioned to take place at the base of the convection zone where the shear is quite strong (the so-called tachocline), e.g. to compensate for the weakness of the interface dynamo \citep{Dikpati05}. In particular,  quenching by shear should be incorporated when assessing the efficiency of dynamo, e.g. the dynamo number given by $D= \alpha \Omega L^3 / (\eta+\beta)^2$, where $\Omega$ is the differential rotation (corresponding to flow shear: $\Omega=\A$ in this paper) and $L$ is a characteristic scale of the system. While it is conventionally thought that the dynamo efficiency increases proportionally to shear  \citep{Kulsrud99} for an $\alpha\Omega$ dynamo, our result suggests that the relation between the dynamo efficiency and the shearing rate is unlikely to be so simple. For instance, in the case of the $\alpha\Omega$ dynamo, the dynamo number $D$ becomes:
\EQ
D = \alpha \A L^3 / (\eta+\beta)^2 \propto \A^{4} \; , 
\EN
which increases with shear much faster than what has been conventionally thought. In the case of an $\alpha^2$ dynamo, we obtain a different scaling:
\EQ
D = \alpha^2 L^4 / (\eta+\beta)^2 \propto \A^{4/3} \; . 
\EN
In the case of $\alpha^2$-dynamo, it is also interesting to examine how the growth rate of the magnetic field scales with shear: using standard formula for the maximum growth rate \citep[see][for instance]{Moffatt78}, we obtain the estimate $\sigma \propto \alpha^2 / \beta \propto \A^{-1}$. 

It is interesting to note that our results are very different from the recent works by \citet{Yousef08} and \citet{Schekochihin08} where the large-scale magnetic field is amplified with a growth-rate scaling as $\A^2$  or $\A$. This is because, in these works, the dynamo instability is triggered by direct interaction between the large scale magnetic field (with both components parallel and perpendicular to the velocity field) and velocity field (i.e. $\langle \BB \rangle \cdot {\bf \nabla} \langle \UU \rangle \neq 0$). Note that in these works, the shear flow is assumed to be weak compared to the diffusion rate, corresponding to our weak shear limit ($\xi \gg 1$). It would be interesting to study the opposite limit of a strong shear ($\xi \ll 1$).  

Finally, we showed that turbulence and transport are enhanced for weak shear while quenched for strong shear. Therefore, there is a critical value of the shear for which the turbulence intensity and transport are maximum. As shown by \citet{Newton07}, this can be due to resonance between the turbulence and shear flow when the characteristic frequency of turbulence matches the advection by shear flow (i.e. the Doppler shifted frequency vanishes).

\begin{acknowledgments}
We would like to thanks A. Brandenburg for valuable comments. E.K. acknowledges the hospitality of Nordita where part of this work was performed. This work was supported by U.K. STFC Grant No. ST/F501796/1.
\end{acknowledgments}

\bibliographystyle{apj}
\bibliography{../../../Biblio/Bib_sun,../../../Biblio/Bib_maths,../../../Biblio/Bib_dynamo,../../../Biblio/Bib_shear,Bib_3DMHD}

\end{document}